\documentclass[aps,showpacs,twocolumn,
superscriptaddress]{revtex4}
\usepackage{graphicx}
\usepackage{dcolumn}
\usepackage{bm}
\usepackage{color}
\usepackage[normalem]{ulem} 
\usepackage[dvipsnames]{xcolor} 
\usepackage{hyperref}
\hypersetup{
  colorlinks=true,        
  linkcolor=blue,         
  citecolor=cyan,         
}
\usepackage{mathrsfs}
\usepackage[utf8]{inputenc}
\usepackage{mathtools}
\usepackage{doi}
\usepackage{amsmath}
\usepackage{amssymb}

\usepackage{scalerel}
\usepackage{tikz}
\usetikzlibrary{svg.path}

\definecolor{orcidlogocol}{HTML}{A6CE39}
\tikzset{
  orcidlogo/.pic={
    \fill[orcidlogocol] svg{M256,128c0,70.7-57.3,128-128,128C57.3,256,0,198.7,0,128C0,57.3,57.3,0,128,0C198.7,0,256,57.3,256,128z};
    \fill[white] svg{M86.3,186.2H70.9V79.1h15.4v48.4V186.2z}
                 svg{M108.9,79.1h41.6c39.6,0,57,28.3,57,53.6c0,27.5-21.5,53.6-56.8,53.6h-41.8V79.1z M124.3,172.4h24.5c34.9,0,42.9-26.5,42.9-39.7c0-21.5-13.7-39.7-43.7-39.7h-23.7V172.4z}
                 svg{M88.7,56.8c0,5.5-4.5,10.1-10.1,10.1c-5.6,0-10.1-4.6-10.1-10.1c0-5.6,4.5-10.1,10.1-10.1C84.2,46.7,88.7,51.3,88.7,56.8z};
  }
}

\newcommand\orcidicon[1]{\href{https://orcid.org/#1}{\mbox{\scalerel*{
\begin{tikzpicture}[yscale=-1,transform shape]
\pic{orcidlogo};
\end{tikzpicture}
}{|}}}}

\begin{document}

\title{Perturbations and quasinormal modes of black holes in general relativity coupled to nonlinear electrodynamics}

\author{Uktamjon Uktamov} \email[Corresponding Author:]{uktam.uktamov11@gmail.com}
\affiliation{School of Physics, Harbin Institute of Technology, Harbin 150001, People’s Republic of China}
\affiliation{Institute for Advanced Studies, New Uzbekistan University,
Movarounnahr str. 1, Tashkent 100000, Uzbekistan}

\author{Bakhtiyor Narzilloev}
	\email{b.narzilloev@newuu.uz}
    \affiliation{Institute for Advanced Studies, New Uzbekistan University, Movarounnahr str. 1, Tashkent 100000, Uzbekistan}

 \author{Ibrar Hussain}	\email{ibrar.hussain@seecs.nust.edu.pk}
    \affiliation{School of Electrical Engineering and Computer Science, National University of Sciences and Technology, H-12, Islamabad, Pakistan}
    \affiliation{Research Center of Astrophysics and Cosmology, Khazar University, 41 Mehseti Street, AZ1096 Baku, Azerbaijan}
    
	\author{Bobomurat Ahmedov}
	\email{ahmedov@astrin.uz}
    \affiliation{Institute for Advanced Studies, New Uzbekistan University, Movarounnahr str. 1, Tashkent 100000, Uzbekistan}
    \affiliation{School of Physics, Harbin Institute of Technology, Harbin 150001, People’s Republic of China}
    \affiliation{Institute of Theoretical Physics, National University of Uzbekistan, Tashkent 100174, Uzbekistan}
\author{Chengxun Yuan}
\email{yuancx@hit.edu.cn}
    \affiliation{School of Physics, Harbin Institute of Technology, Harbin 150001, People’s Republic of China}

\date{\today}

\begin{abstract}
We investigate the quasinormal spectra and time-domain evolution of scalar, electromagnetic, and gravitational perturbations of the Einstein--Bronnikov black hole in Einstein gravity coupled to nonlinear electrodynamics. The effective potentials associated with all perturbation sectors are shown to form positive potential barriers that vanish at the event horizon and spatial infinity. Their height increases with the magnetic charge, while the scalar perturbation possesses the largest potential barrier. Unlike the Reissner--Nordström black hole, axial and polar electromagnetic perturbations are characterized by distinct effective potentials, demonstrating the breaking of isospectrality due to nonlinear electrodynamics. Quasinormal frequencies are computed using the sixth-order WKB approximation and the asymptotic iteration method, yielding excellent agreement for both fundamental and higher overtone modes. The oscillation frequencies increase monotonically with the magnetic charge, whereas the damping rates exhibit only moderate variations, confirming the linear stability of the Einstein--Bronnikov black hole under all perturbations considered. Time-domain profiles display exponentially damped ringdown signals consistent with the frequency-domain analysis. These results demonstrate that nonlinear electrodynamics leaves measurable imprints on black hole perturbations and may provide observational signatures for testing regular black hole geometries through future gravitational wave observations.
\end{abstract}


\maketitle

\section{Introduction}

The observation of gravitational waves (GWs) by the LIGO and Virgo collaborations \cite{LIGOScientific:2016aoc}, produced by the merger of supermassive black holes represents one of the most profound confirmations of Einstein’s theory of General Relativity (GR). The observations of the GWs have opened a new window for researchers to probe gravity when the spacetime curvature is extremely large and relativistic effects are dominant. The GW signals emitted during the merger of supermassive black holes and the subsequent relaxation of black holes carry very useful information about the geometry of spacetime and the physical properties of the resulting compact objects \cite{Yunes:2016jcc}. The remarkable agreement between the observed GW signals and the predictions of GR has provided stringent tests of the theory in a regime that is difficult to explore through conventional astronomical observations \cite{LIGOScientific:2019fpa}. Consequently, GW astronomy has established black hole perturbations and ringdown as powerful tools for investigating the fundamental properties of gravity in strong field regime.

As a result of the merger of black holes, the newly formed black hole undergoes the ringdown phase, during which perturbations of the spacetime decay through a superposition of damped oscillations known as quasinormal modes (QNMs) in the literature \cite{Berti:2009kk}. The complex QNM frequencies consist of two parts, oscillation frequencies and damping rates, and are determined by the background spacetime and its physical parameters \cite{Redondo-Yuste:2023ipg}. Interestingly, the spectrum of QNMs provides a characteristic fingerprint of black holes, which can be used to probe the mass and angular momentum of black holes \cite{Berti:2025hly}. It may be also useful to understand the possible deviations from the predictions of GR by studing the QNMs in different gravitational theories. The QNMs were first studied in the context of Schwarzschild black hole \cite{Kokkotas:1999bd} and subsequently extended to the Kerr black hole \cite{Yang:2012he}, where perturbations are governed by the Regge–Wheeler, Zerilli, and Teukolsky formalisms \cite{Pound:2021qin}. Since then, a variety of analytical and numerical techniques, including the WKB approximation, continued-fraction method \cite{Daghigh:2022uws}, asymptotic iteration method (AIM) \cite{Cho:2011sf}, time-domain integration \cite{Dubinsky:2024gwo}, and spectral approaches \cite{Chung:2023zdq}, have been developed to study QNM spectra with improving accuracy.

QNMs have been extensively investigated for a wide range of black hole spacetimes \cite{Dreyer:2003bv,Bhattacharyya:2018qbe,Silva:2024ffz,Manfredi:2017xcv,Blazquez-Salcedo:2018pxo,Zhang:2025dzt} in GR and modified or alternative gravitational theories \cite{Chen:2021cts,Shankaranarayanan:2022wbx,Davlataliev:2024mjl,Bojowald:2024lhr,DiRusso:2025qpf}. Initially researchers focused primarily on the Schwarzschild and Kerr geometries to establish the fundamental properties of their perturbation spectra. In the subsequent investigations the Reissner–Nordström black hole \cite{Andersson:2003fh,UktamjonUktamov:2025aqz}, higher-dimensional black holes \cite{Zhidenko:2006rs}, and black holes surrounded by matter fields \cite{C:2024cnk,Uktamov:2025bth}, dark matter \cite{Jusufi:2019ltj,UktamjonUktamov:2025qts}, or other exotic structures \cite{Atamurotov:2022nim,MoraisGraca:2017hrf,UktamjonUktamov:2025cso}, are considered. In recent studies, QNMs have also been considered in modified theories of gravity, including the Einstein–Gauss–Bonnet gravity \cite{Gera:2023dld}, $f(R)$ gravity \cite{Karmakar:2024xwr}, scalar–tensor theories of gravity \cite{Franciolini:2018uyq}, Horndeski gravity \cite{Tattersall:2018nve}, and various quantum-corrected and regular black hole models \cite{Moreira:2023cxy,Gong:2023ghh,Lutfuoglu:2025hwh}. These studies on the QNM of different modified theories of gravity have demonstrated that the additional parameters of these theories can modify the real and as well the imaginary parts of the QNM frequencies, thereby leaving potentially observable imprints on the ringdown signal. Particularly, the dependence of QNM frequencies on the charge and spin of black holes, the coupling constants of gravitational theories, and additional geometric or matter parameters makes them a useful theoretical tool for distinguishing between different black hole models and testing the validity of GR in the strong field regime \cite{Yagi:2016jml,Lagos:2024ekd}.

The QNM spectrum is therefore of particular interest in the context of GW observations, since the dominant ringdown modes may be extracted from sufficiently strong signals and compared with theoretical predictions. Such comparisons provide a direct avenue for testing the black hole no-hair hypothesis and searching for signatures of new physics beyond GR. Motivated by these developments, the investigation of QNMs in alternative black-hole geometries has become an active area of research \cite{Chrysostomou:2022evl}. In this work, we study the QNM spectrum of a black hole in GR coupled to nonlinear electrodynamics, with particular emphasis on the effects of the oscillation frequencies and damping times. 

In particular, we investigate the quasinormal (QN) spectra and time-domain evolution of scalar, electromagnetic, and gravitational perturbations of the Einstein-Bronnikov (EB) black hole, a regular black hole solution arising in Einstein gravity coupled to nonlinear electrodynamics (NLED) \cite{Bronnikov:2000vy}. Regular black holes supported by NLED offer an interesting setting in which deviations from the standard Reissner-Nordström geometry can leave characteristic imprints on the ringdown signals. We show that the effective potentials governing all three perturbation sectors form positive potential barriers that vanish at both the event horizon and spatial infinity. The height of these barriers increases with the magnetic charge, with the scalar perturbation exhibiting the largest barrier among the three sectors. Unlike the Reissner-Nordström case, the axial and polar electromagnetic perturbations of the EB black hole are governed by distinct effective potentials, indicating a breaking of isospectrality induced by the nonlinear electrodynamic coupling. The corresponding QNM frequencies are calculated using the sixth-order WKB approximation and the AIM \cite{Cho:2009cj}, with excellent agreement obtained between the two approaches for both fundamental and higher-overtone modes. We find that the oscillation frequencies increase monotonically with the magnetic charge, while the damping rates undergo only moderate variations. The positivity of the effective potentials, together with the resulting QNM behavior, indicates that the EB black hole is linearly stable against all perturbations considered. The time-domain evolution further confirms this picture, exhibiting exponentially damped ringdown signals in agreement with the frequency-domain analysis. Our results demonstrate that NLED can produce distinct and potentially observable modifications in black hole perturbation spectra, particularly through the breaking of electromagnetic isospectrality and the dependence of QNM frequencies on the magnetic charge. These features highlight the potential of black hole ringdown observations to distinguish regular black hole geometries from their general relativistic counterparts and to constrain possible modifications of the matter sector in the strong field regime.

This work is organized as follows. In the next section, we briefly discuss the EB black hole. In Sec. III, we study the behavior of the EB black hole under gravitational perturbations. In the same section, we investigate the response of scalar and electromagnetic perturbations in the EB black hole spacetime. In Sec. IV, we calculate the QN frequencies associated with gravitational, scalar, and electromagnetic perturbations. In Sec. V, we investigate the time-domain evolution of scalar, axial and polar electromagnetic, and gravitational perturbations of the EB black hole in the framework of NLED. Finally, we summarize our main findings and present our conclusions in Sec. VI.

\section{Einstein-Bronnikov black hole}

The action for the spacetime of the EB black hole can be expressed as (\cite{Bronnikov:2000vy,Gao:2025lir}):
\begin{eqnarray}\label{eq.}
  S=\frac{1}{16\pi}\int d^4x\sqrt{-g}\left[R-L(F)\right]\,,  
\end{eqnarray}
here $R$ represents the Ricci scalar, $g$ denotes the determinant of the metric tensor, $L(F)=F\text{cosh}^{-2}\left[\frac{q_m^{3/2}}{2M}(\frac{F}{2})^{1/4}\right]$ is the Lagrangian density associated with Einstein–Born–Infeld (EBI) NLED theory, $q_m$ is the magnetic charge. 

The covariant equations of motion are:
\begin{subequations}\label{eq.Einstein}
    \begin{align}
    &G_{\mu\nu}=8\pi T_{\mu\nu}=\left[2L_FF_{\rho\mu}F_\nu^\rho-\frac{1}{2}g_{\mu\nu}L(F)\right]\,,\\\label{eq.equation of motion}
&\nabla_\mu(L_FF^{\mu\nu})=0\,,
    \end{align}
\end{subequations}
in which $L_F=\partial_FL$, $F_{\mu\nu}=\partial_\mu A_\nu-\partial_\nu A_\mu$ is the electromagnetic field. 

The line element of the EB black hole  can be expressed as:
\begin{subequations}\label{eq.the metric}
    \begin{align}
        &ds^2=-f(r)dt^2+\frac{dr^2}{f(r)}+r^2d\Omega^2\,,\\ 
        &f(r)=1-\frac{2M}{r}\left[1-\text{tanh}(\frac{q_m^2}{2Mr})\right]\,,
    \end{align}
\end{subequations}
and the electromagnetic four potential is:
\begin{eqnarray}\label{eq.A}
\overline{A}_\mu=-q_m\cos{\theta}\delta_\mu^\phi\,. 
\end{eqnarray}

\begin{figure*}[t]
\includegraphics[width=0.45\textwidth]{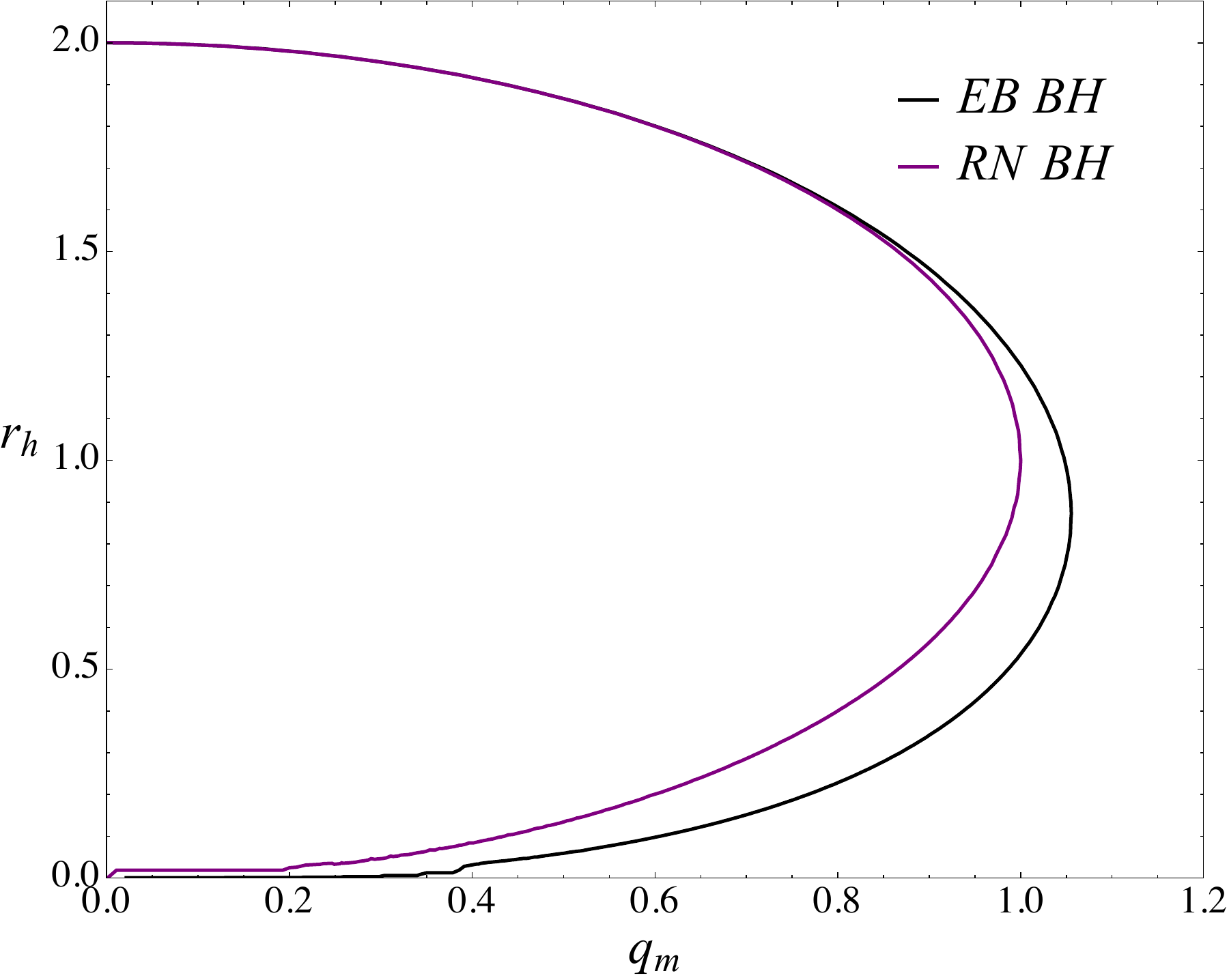}
\caption{Dependence of the horizon radius $r_h$  on magnetic charge $q_m$ for RN black hole and EB black hole.}
    \label{fig. rh}
\end{figure*}

Subsequently, the event horizon $r_h$ of the EB black hole can be found using condition $f(r)=0$, so we give dependence of the $r_h$ on the magnetic charge $q_m$ in Fig.(\ref{fig. rh}). One can see from Fig.(\ref{fig. rh}) that EB black holes have two inner and outer horizons which are coincides in the extreme value of the magnetic charge $q_m^*$ and we compare event horizons of the Reissner-Nordstr\"om (RN) black holes with the event horizons of the EB black holes in Fig.(\ref{fig. rh}). The extreme value of the magnetic charge $q_m^*$ and corresponding event horizon $r_h^*$ can be found solving $f(r)=\frac{\partial f}{\partial r}=0$ equations simultaneously, as $q_m^*=1.05539$ and $r_h^*=0.871247$. 

\section{Perturbation}
We devote this section to studying the behavior of the EB black hole under gravitational perturbations. Additionally, we study the response of scalar and electromagnetic configurations to perturbations in the black hole spacetime. We now proceed to analyze these three distinct classes of perturbations separately.

\subsection{Scalar perturbation}
We here consider the perturbation of the massless scalar field in the spacetime of the EB black hole. The behavior of the scalar field $\Phi$ is described by the Klein–Gordon equation, $\square \Phi=0$. Introducing a perturbation to the scalar field as $\Psi(t,r,\theta,\phi)+\delta\Psi(t,r,\theta,\phi)$, the perturbation term $\delta\Psi(t,r,\theta,\phi)$ must itself satisfy the covariant equation (\cite{Ovgun:2025ehi}):
\begin{eqnarray}\label{eq.deltaPhi}
    \frac{1}{\sqrt{-g}}\partial_\alpha\left(\sqrt{-g}g^{\alpha\beta}\partial_\beta\delta\Psi\right)=0\,.
\end{eqnarray}
Here $g$ is the determinant of the metric (\ref{eq.the metric}). To separate the angular part of the Klein–Gordon equation (\ref{eq.deltaPhi}), we expand the scalar field in spherical harmonics:
\begin{eqnarray}\label{eq.deltaPhi2}
\delta\Psi(t,r,\theta,\phi)=\frac{\Phi_{sc.}(t,r)}{r}Y_{l}(\theta,\phi)\,,
\end{eqnarray}
where $Y_l$ is a scalar spherical harmonic function of degree $l$. Solving Eq.(\ref{eq.deltaPhi}) yields:
\begin{eqnarray}\label{eq.Psi_sc}
    \frac{\partial^2\Phi_{sc}}{\partial r_*^2}-\frac{\partial^2\Phi_{sc.}}{\partial t^2}-V_{sc}\Phi_{sc}=0\,,
\end{eqnarray}
in which
\begin{eqnarray}\label{eq.Vsc}
    \frac{dr_*}{dr}=\frac{1}{f(r)}\,,\,\,\,\,V_{sc}=f(r)\left[\frac{l(l+1)}{r^2}+\frac{f(r)_{,r}}{r}\right]\,,
\end{eqnarray}
where $r$ in the subscript denotes derivative with respect to $r$ and hereafter we use the property of the spherical harmonic function $\cot{\theta}Y_l(\theta,\phi)_{,\theta}+Y_l(\theta,\phi)_{,\theta\theta}=-l(l+1)Y_l(\theta,\phi)$. 

Also electromagnetic perturbations of black holes within the framework of NLED can be analyzed  by incorporating the electromagnetic perturbations directly into the gauge potential (\cite{Toshmatov:2018tyo,Uktamjon:2024fjb}) as:
\begin{eqnarray}\label{eq.A}
A_\mu=\overline{A}_\mu+\delta A_\mu\,,
\end{eqnarray}
where electromagnetic perturbation can be divided into axial $\delta A_\mu^{ax.}$ and polar part $\delta A_\mu^{pol.}$ so total electromagnetic perturbation is $\delta A_\mu=\delta A_\mu^{ax.}+\delta A_\mu^{pol.}$. Hence, we will study axial and polar electromagnetic perturbations in following subsections.

\subsection{Axial Electromagnetic perturbations of EB black holes}
Axial $\delta A^{ax.}_\mu$ electromagnetic perturbations can be incorporated to the gauge potential (\ref{eq.A}) as
\[
\delta A^{ax.}_\mu = \sum_{\ell, m} 
\begin{pmatrix}
0 \\[4pt]
0 \\[4pt]
\dfrac{a^{\ell m}_{em}(t, r) \partial_\phi Y_{\ell m}(\theta, \phi)}{\sin \theta} \\[8pt]
-a^{\ell m}_{em}(t, r) \sin \theta \, \partial_\theta Y_{\ell m}(\theta, \phi)
\end{pmatrix}.
\]\,

Consequently, electromagnetic field strength $F=F_{\mu\nu}F^{\mu\nu}$ can be found up to first order perturbation terms as:
\begin{eqnarray}\label{eq.F}
   F\approx\frac{2q_m}{r^4}+\frac{4q_ml(l+1)a^{\ell m}_{em}(t,r)Y_{lm}(\theta,\phi)}{r^4}\,, 
\end{eqnarray}
here  the first term represents the field strength of the unperturbed electromagnetic background, $\bar{F}$, whereas the second term denotes the perturbation contribution, $\delta F$, so that the total field strength is given by $F = \bar{F} + \delta F$.
The $L_F$ can be expressed as:
\begin{eqnarray}
    L_F&=&\overline{L}_F+\overline{L}_{FF}\delta F=\\\nonumber
    &=&\overline{L}_F+\overline{L}_{FF}\frac{4q_ml(l+1)a^{\ell m}_{em}Y_{lm}}{r^4}\,.
\end{eqnarray}
Solving equation of motion (2b) gives expression (\cite{Toshmatov:2018tyo}) up to first order perturbation:
\begin{eqnarray}\label{eq.Field eq.}
    a^{\ell m}_{em,tt}+\left(\frac{f^2(r)}{\alpha}a^{\ell m}_{em,r}\right)_{,r}+f(r)\frac{l(l+1)}{r^2}\gamma a^{\ell m}_{em}=0\,,
\end{eqnarray}
where new variables are given in Appendix (\ref{app}). Now introducing  tortoise coordinate and choosing electromagnetic field $\Phi^{ax.}_{em}$ as:
\begin{subequations}
    \begin{align}
        &\frac{dr_*}{dr}=\frac{1}{f(r)}\,,\\
    &\Phi^{ax.}_{em}=\sqrt{\frac{f(r)}{\alpha}}a^{\ell m}_{em}\,,
    \end{align}
\end{subequations}
yields to the wave equation:
\begin{eqnarray}\label{eq.wave}
    -\frac{\partial \Phi^{ax.}_{em}}{\partial t^2}+\frac{\partial\Phi^{ax.}_{em}}{\partial r_*}-V_{em}^{ax.}\Phi^{ax.}_{em}=0\,,
\end{eqnarray}
in which
\begin{eqnarray}\label{eq.Vem}
    V_{em}^{ax.}&=&f\frac{l(l+1)}{r^2}\gamma-\frac{\alpha^2}{4}\left[(\frac{f(r)}{\alpha})_{,r}\right]^2+\\\nonumber
&+&\alpha\left[f(r)(\frac{f(r)}{\alpha})_{,r}\right]_{,r}\,.
\end{eqnarray}

\subsection{Polar Electromagnetic perturbations of the EB black holes}

Now our aim is to analyze polar electromagnetic perturbation of EB black holes by introducing polar electromagnetic perturbation into gauge potential (\ref{eq.A}) as (\cite{2018PhRvD..98h5021T}):

\[
\delta A^{pol.}_\mu = \sum_{\ell, m} 
\begin{pmatrix}
d^{\ell m}(t,r)Y_{\ell m}(\theta,\phi) \\[4pt]
h^{\ell m}(t,r)Y_{\ell m}(\theta,\phi) \\[4pt]
k^{\ell m}(t,r)\partial_\theta Y_{\ell m}(\theta,\phi) \\[4pt]
k^{\ell m}(t,r)\partial_\phi Y_{\ell m}(\theta,\phi)
\end{pmatrix}.
\]\,

Again, electromagnetic field strength can be approximated up to first order perturbation term:
\begin{eqnarray}
    F\approx\frac{2q_m^2}{r^4}\,,
\end{eqnarray}
which is the same as for the Reissner--Nordstr\"om black hole. Therefore, the Lagrangian density $L$ of the EB black hole also remains unchanged. 

Now using Eq.(\ref{eq.equation of motion}) and introducing new variable as:
\begin{eqnarray}\label{eq.Phi pol}
    \Phi^{pol.}_{em}(t,\,r)=\frac{1}{r^2\sqrt{L(F)}}\left(\partial_th^{\ell m}-\partial_rd^{\ell m}\right)\,,
\end{eqnarray}
we arrive at the well-known wave equation as (see (\cite{2018PhRvD..98h5021T}) for details):
\begin{eqnarray}\label{eq.Vave pol}
-\frac{\partial \Phi^{pol.}_{em}}{\partial t^2}+\frac{\partial\Phi^{pol.}_{em}}{\partial r_*}-V_{em}^{pol.}\Phi^{pol.}_{em}=0\,,
\end{eqnarray}
in which
\begin{eqnarray}\label{eq.Vpolar}
    V_{em}^{pol.}=f\frac{l(l+1)}{r^2}-\frac{\alpha}{4}\left((\frac{f}{\alpha})_{,r}-\left[f(\frac{f}{\alpha})_{,r}\right]_{,r}\right)\,.
\end{eqnarray}

\subsection{Gravitational and Electromagnetic perturbation}

In this section, we study both electromagnetic and gravitational perturbation simultaneously. Now, the electromagnetic perturbation of the EB black hole can be expressed as (\cite{2019PhRvD..99f4043T}):
\begin{eqnarray}\label{eq.A gr}
\delta A^{ax.}_\mu = \sum_{\ell, m} 
\begin{pmatrix}\label{eq.A gr}
0 \\[4pt]
0 \\[4pt]
0 \\[4pt]
\psi(t,r)\sin{\theta}\partial_\theta P_k(\cos{\theta})\,
\end{pmatrix}
\end{eqnarray}
where $P_k(\cos{\theta})$ is the Legendre polynomials. Also, the gravitational perturbation in the ``Regge-Wheeler" gauge can be expressed as $g_{\mu\nu}=\overline{g}_{\mu\nu}+h_{\mu\nu}$ (\cite{2019PhRvD..99f4043T}) :
\begin{eqnarray}\label{eq.h}
h_{\mu\nu} = 
\begin{pmatrix}
0 & 0 & 0 & \epsilon\,h_0(t,\,r) \\
\ast & 0 & 0 & \epsilon\,h_1(t,\,r) \\
\ast & \ast & 0 & 0 \\
\ast & \ast & \ast & 0
\end{pmatrix}
 \, \sin\theta \, \partial_\theta P_\ell(\cos\theta)\,,
\end{eqnarray}
here  $*$ denotes the symmetric components and $\epsilon=1$ denotes gravitational perturbation up to first order. Now inserting Eqs.(\ref{eq.A gr},~ \ref{eq.h}) into equation of the motion (\ref{eq.Einstein}), yields the following system of equations:
\begin{subequations}
    \begin{align}
        & h_{0,rr}-\frac{2}{r}h_{1,t}-h_{1,tr}-\frac{h_0}{r^2f}\left[\lambda+2f+(r^2f_{,r})_{,r}+Lr^2\right]=0\,,\\\label{eq.2 d}
        &h_{0,tr}-h_{1,tt}-\frac{2}{r}h_{0,t}-\frac{h_1f}{r^2}\left[\lambda+(r^2f_{,r})_{,r}+Lr^2\right]=0\,,\\\label{eq.3 d}
        & f(h_1f)_{,r}-h_{0,t}=0\,,
    \end{align}
\end{subequations}
where we have used Legendre polynomials property $\sin^2{\theta}P_{k,\theta\theta}(\cos{\theta})-2\cos{\theta}P_{k,\theta}(\cos{\theta})=-l(l+1)P_k(\cos{\theta})$ and introduced new variable as $\lambda=(l+2)(l-1)$.  After finding $h_{0,t}$ from Eq.(\ref{eq.3 d}) and putting it into Eq.(\ref{eq.2 d}) paves a way to find second order differential equation:
\begin{eqnarray}\label{eq.2 order}
    &\frac{h_{1,tt}}{f^2}-h_{1,rr}-\left(\frac{3f_{,r}}{f}-\frac{2}{r}\right)h_{1,r}\\\nonumber
    &-\frac{h_1f}{r^2}\left[\lambda+(r^2f_{,r})_{,r}+Lr^2\right]=0\,.
\end{eqnarray}

\begin{figure*}[t]\centering
\includegraphics[width=0.3\textwidth]{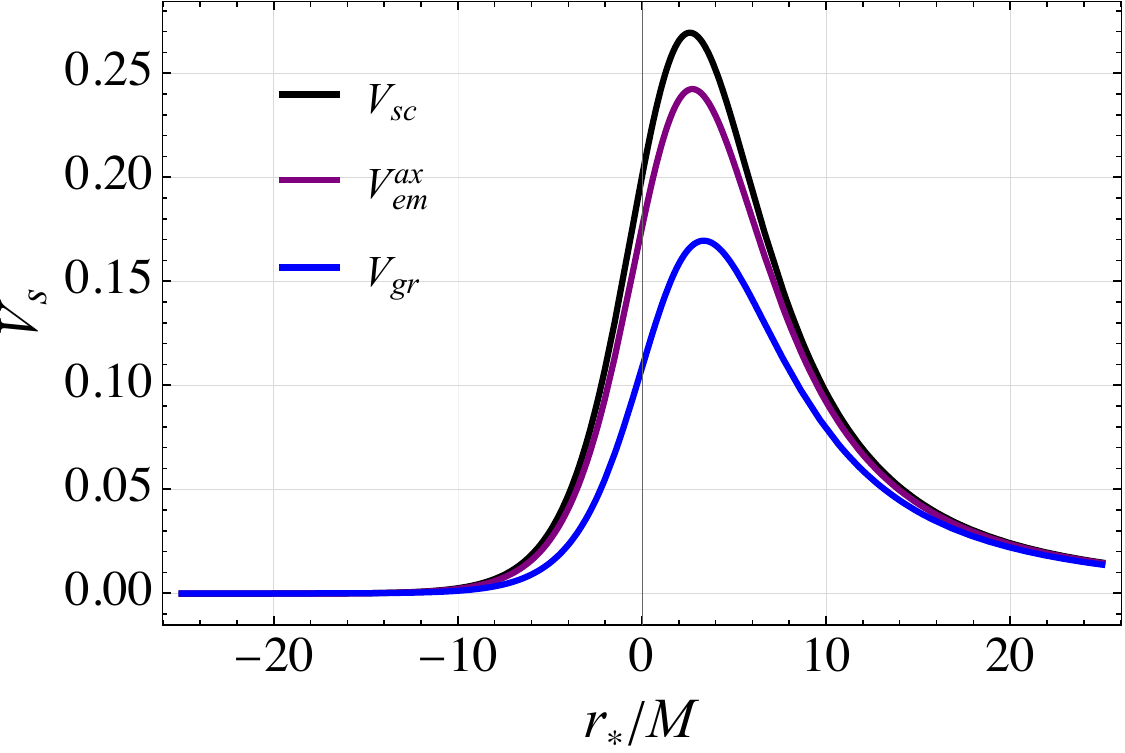}
\includegraphics[width=0.31\textwidth]{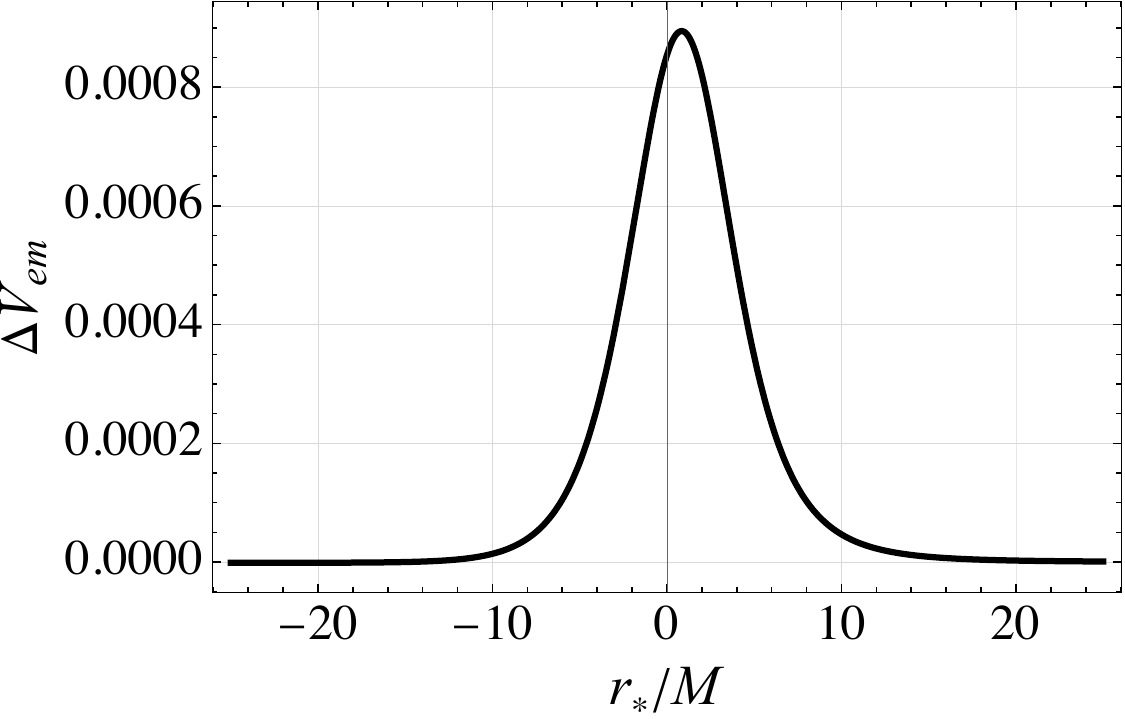}
\includegraphics[width=0.3\textwidth]{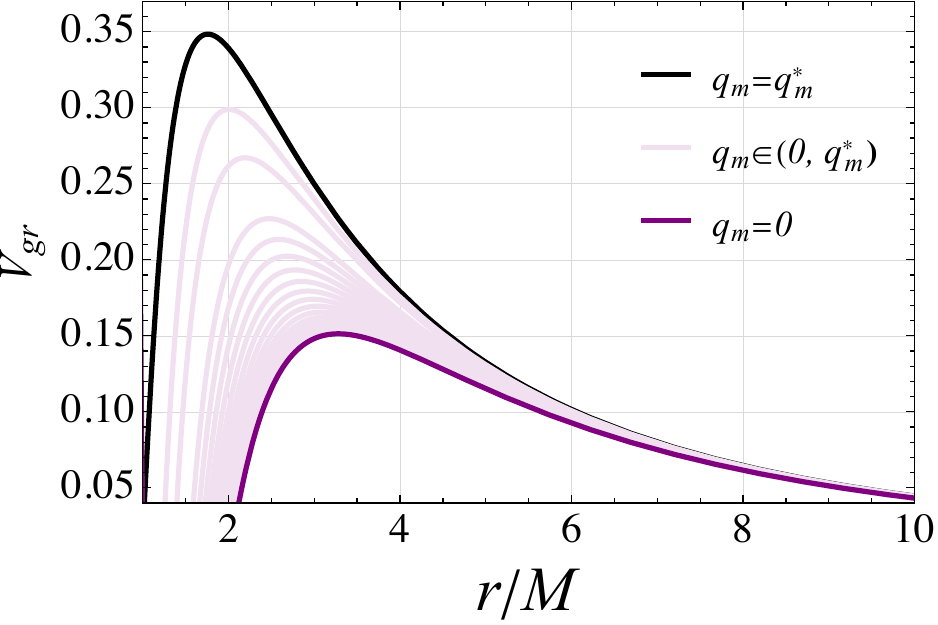}
\caption{The dependence of the effective potential for the three types of perturbations $V_s$ (the left panel), $\Delta V_{em}=V_{em}^{pol}-V_{em}^{ax}$ as a function tortoise coordinate (the middle panel). The right panel shows the radial profile of the effective potential for gravitational perturbations of an EB black hole, where the magnetic charge increases in increments of 0.05. The parameters are set to  $l=2$, $q_m=0.5$.}
    \label{fig. V}
\end{figure*}

Subsequently, introducing new function as:
\begin{eqnarray}
    \Phi_{gr}(t,r)=\frac{f}{r}h_1(t,r)\,,
\end{eqnarray}
we arrive the wave equation as:
\begin{eqnarray}\label{eq.Schr.}
    -\frac{\partial^2\Phi_{gr}}{\partial t^2}+\frac{\partial\Phi_{gr}}{\partial r_*}-V_{gr}\Phi_{gr}=0\,,
\end{eqnarray}

in which
\begin{eqnarray}\label{eq.Vgr}
    V_{gr}=f\left[\frac{l(l+1)}{r^2}+\frac{r(rf_{,r})_{,r}+2(f-1)}{r^2}+L(F)\right]\,.
\end{eqnarray}
Harmonically time dependent perturbations:
\begin{eqnarray}\label{eq.harm}
    \Phi_{s}(t,r)=\Psi_s(r)e^{-i\omega_st}\,,\,\,\,\,\,(s=\text{sc, em, gr})\,,
\end{eqnarray}
converts the master wave Eqs.(\ref{eq.Psi_sc},\ref{eq.wave},\ref{eq.Vave pol},\ref{eq.Schr.}) into Schr\"{o}dinger-like wave equations with scattering potential:
\begin{eqnarray}\label{eq.wave1}
    \frac{\partial^2\Psi_s(t,r)}{\partial r^2_*}+\left(\omega_s^2-V_s\right)\Psi_s(t,r)=0\,,
\end{eqnarray}
here $\omega$ corresponds to the QN frequency. Asymptotic flatness guarantees that the effective potentials $V_s$ vanish at the event horizon and tend to zero at infinity. Also, the plots in Fig.(\ref{fig. V}) reveals further characteristic of the effective potential $V_s$. It is clear from graphics in Fig.(\ref{fig. V}) that potential barriers increase with growing the value of the magnetic charge $q_m$ of the EB black hole and the highest value of the effective potential corresponds to the scalar perturbation.

\section{QUASINORMAL MODES}

In this section, we compute the QN frequencies for gravitational, scalar, and electromagnetic perturbations.  As $\omega$ is a complex quantity comprising real and imaginary parts, its computation cannot be analytically. To address this, we adopt established (semi) analytic approximation methods, including the WKB and AIM, in conjunction with numerical approaches. 

\subsection{WKB method}
In this section we calculate QNM using powerful semi analytical WKB approach. To ensure physical consistency of the solution, the waveform should demonstrate purely ingoing behavior at the event horizon and purely outgoing behavior at spatial infinity, as follows:
\[
\psi(r) =
\begin{cases} 
e^{-i\omega r_*} \text{ at } r_* \to -\infty, & (r \to r_{\text{h}}), \\
e^{i\omega r_*} \text{ at } r_* \to \infty, & (r \to \infty).
\end{cases}
\]
Firstly, Schutz and Will (\cite{1985ApJ...291L..33S}) applied the WKB method to calculate QNMs. Then Iyer and Will extended the method to third order (\cite{PhysRevD.35.3621}), Konoplya extended to sixth order (\cite{Konoplya:2003ii}),  and recently the method was extended to thirteenth order by Matyjasek and Opala (\cite{Matyjasek:2017psv}) as
\begin{eqnarray}
    \omega^2&=&V_0+A_2(\mathcal{K}^2)+A_4(\mathcal{K}^2)+...\\\nonumber
    &-&i\mathcal{K}\sqrt{-2V_2}\left[1+A_3(\mathcal{K}^2)+A_5(\mathcal{K}^2)+...\right]\,,
\end{eqnarray}
where $V_0$ is the peak of the effective potential at $r_0$, $\mathcal{K}$ are half-integer quantities, and $A_k(\mathcal{K}^2)$ are order-$k$ corrections given by polynomials in $\mathcal{K}^2$ featuring rational coefficients.
 
The WKB method is very accurate when the potential barrier (\ref{fig. V}) vary slowly compared to the wavelength of the wave so the method is effective for wide potential barrier. However, as the charge $q_m$ of the EB black hole increases, the effective potential narrows (see right panel of Fig.(\ref{fig. V})), which reduces the accuracy of the WKB method, necessitating the use of additional methods alongside it to cross-verify and obtain more precise QN frequencies.

\subsection{Asymptotic iteration method}
Now we calculate QNM using AIM (\cite{Toshmatov:2025rln,Mamani:2022akq,Cho:2009cj}). The wave equation given in Eq. (\ref{eq.wave1}) can be expressed as:
\begin{eqnarray}\label{eq.wave radial}
    \Psi_{s,rr}(t,r)+\frac{f(r)_{,r}}{f(r)}\Psi_{s,r}(t,r)+\frac{\omega^2_s-V_s}{f(r)^2}\Psi_{s}(t,r)=0\,.
\end{eqnarray}

The metric function (\ref{eq.the metric}) can also be factorized in terms of its zeros as:
\begin{eqnarray}\label{eq.lapse with 0}
    f(r)=\frac{(r-r_1)(r-r_h)}{r^2}\,,
\end{eqnarray}
where $r_1$ and $r_h$ denote the inner and event horizons of the EB black holes, respectively.

Subsequently, introducing new variable as $u=1/r$ takes wave equation (\ref{eq.wave radial}) into following form:
\begin{widetext}
\begin{eqnarray}\label{eq.the wave with u}
    \Psi_{sc,uu}+\left(\frac{2}{u}-\frac{f_{,u}}{f}\right)\Psi_{sc,u}+\left(\frac{\omega^2_{sc}}{u^4f^2}+\frac{uf'-l(l+1)}{u^2f}\right)\Psi_{sc}=0\,,
\end{eqnarray}
\end{widetext}
hereafter, we outline the AIM for the scalar perturbation case (s=sc) of the EB black hole, described by Eq. (\ref{eq.Psi_sc}). In order to avoid the divergence of the wave function at spatial infinity  ($r\to\infty$ or $u\to0$), we express the wave function as:
\begin{eqnarray}\label{eq.Psi 1}
    \Psi=\exp{(i\omega r_*)}\chi(u)\,.
\end{eqnarray}
Now Eq.(\ref{eq.the wave with u}) is simplified as:
\begin{eqnarray}\label{eq.the wave 2}
    \chi_{,uu}+A\chi_{,u}+B\chi=0\,,
\end{eqnarray}
in which:
\begin{widetext}
    \begin{subequations}
        \begin{align}
            &A=\frac{u\left[2-u(r_1+r_h)\right]-2i\omega}{u^2(r_1u-1)(r_hu-1)}\,,\\
            &B=-\frac{l(l+1)+u(r_1+r_h-2r_1r_hu)}{u^2(r_1u-1)(r_hu-1)}-\frac{2i\omega(r_1+r_h-2r_1r_hu)}{u^2(r_1u-1)^2(r_hu-1)^2}\,.
        \end{align}
    \end{subequations}
\end{widetext}
Imposing the appropriate boundary condition at the black hole horizon  $r_h$, we express the function $\chi$ as:
\begin{eqnarray}\label{eq.chi}
    \chi(u)=\left(u-\frac{1}{r_h}\right)^{-\frac{i\omega}{\kappa_h}}\xi(u)\,,
\end{eqnarray}
in which $\kappa_h$ represents the surface gravity of the black hole horizon:
\begin{eqnarray}\label{eq.kappa}
    \kappa_h=\frac{r_h-r_1}{2r_h^2}\,.
\end{eqnarray}

Finally, substituting Eqs.(\ref{eq.chi},\ref{eq.kappa}) into the wave Eq.(\ref{eq.the wave 2}) yields to the wave function as:
\begin{eqnarray}\label{eq.the wave 3}
    \xi_{,uu}=\lambda_0\xi_{,u}+s_0\xi\,,
\end{eqnarray}
with new coefficients
\begin{widetext}
    \begin{subequations}
        \begin{align}
            &\lambda_0=\frac{u\left[-2+u(r_1+r_h)\right]+2i\omega}{u^2(r_1u-1)(r_hu-1)}-\frac{4ir_h^3\omega}{(r_1-r_h)(r_hu-1)}\,,\\
            &s_0=\frac{2ir_h^3\omega\left[u(-2+u(r_1+r_h))+2i\omega\right]}{u^2(r_1-r_h)(r_1u-1)(r_hu-1)^2}+\frac{2ir_h^4\omega\left[r_1-r_h(1+2ir_h\omega)\right]}{(r_1-r_h)^2(r_hu-1)^2}\\\nonumber
            &+\frac{l(l+1)+u(r_1+r_h-2r_1r_hu)}{u^2(r_1u-1)(r_hu-1)}+\frac{2i\omega(r_1+r_h-2r_1r_hu)}{u^2(r_1u-1)^2(r_hu-1)^2}\,.
        \end{align}
    \end{subequations}
\end{widetext}

After differentiating equation (\ref{eq.the wave 3}) $n$ times with respect to $u$, we obtain the following equation (\cite{Cho:2009cj}):
\begin{eqnarray}\label{eq.the wave 4}
\xi^{(n+2)}=\lambda_n\xi_{,u}+s_nu\,,    
\end{eqnarray}
in which the coefficients fulfill the following recurrence relations:
\begin{subequations}\label{eq. lambda}
    \begin{align}
        &\lambda_n=\lambda_{n-1,u}+s_{n-1}+\lambda_0\lambda_{n-1}\,,\\
        &s_n=s_{n-1,u}+s_0\lambda_{n-1}.
    \end{align}
\end{subequations}
Subsequently, the coefficients $\lambda_n$ and $s_n$ are written as Taylor series expansions about an arbitrary point $u_0$, given by:
\begin{subequations}\label{eq.lambda 1}
    \begin{align}
        &\lambda_n=\sum_{i=0}^{\infty}c_n^i(u-u_0)^i\,,\\
        &s_n=\sum_{i=0}^{\infty}d_n^i(u-u_0)^i\,,
    \end{align}
\end{subequations}
with $c_n^i$ and $d_n^i$ representing the $i$th-order Taylor coefficients of $\lambda_n$ and $s_n$, respectively. Using Eqs.(\ref{eq. lambda},\ref{eq.lambda 1}) we can find expressions for the $c_n^i$ and $d_n^i$ as:
\begin{subequations}
    \begin{align}
        &c_n^i=(i+1)c_{n-1}^{i+1}+d_{n-1}^i+\sum_{k=0}^ic_0^kc_{n-1}^{i-k}\,,\\
        &d_n^i=(i+1)d_{n-1}^{i+1}+\sum_{k=0}^id_0^kc_{n-1}^{i-k}\,,
    \end{align}
\end{subequations}
which paves a way to find recurrence relation as:
\begin{eqnarray}\label{eq.rec}
    d_n^0c_{n-1}^0-c_n^0d_{n-1}^0=0\,,
\end{eqnarray}
here we have used the asymptotic behavior of the coefficients for the large values of the $n$ which is $\frac{s_n}{\lambda_n}=\frac{s_{n-1}}{\lambda_{n-1}}$. 
Finally, by evaluating $\lambda_0$ and $s_0$ and selecting an appropriate iteration number $n$, one can determine the QNMs of the perturbations.

\begin{table}[ht!]
    \centering
    \begin{tabular}{|l|c|c|r|}
     \hline
       $l$ & $q_m[M]$ & WKB & AIM  \\
    \hline
      2   & 0   & 0.48364 - 0.09677 i   & 0.48364 - 0.09676 i  \\
         &    & 0.45759 - 0.09501 i & 0.45759 - 0.09500 i \\
         &    & 0.45759 - 0.09501 i & 0.45759 - 0.09500 i \\
      &    & 0.37362 - 0.08889 i & 0.37367 - 0.08896 i \\
      &    &  &   \\
         & 0.2   & 0.48693 - 0.09697 i & 0.48693 - 0.09697 i  \\
         &    & 0.46983 - 0.09151 i & 0.46983 - 0.09151 i  \\
         &    & 0.46081 - 0.09523 i & 0.46083 - 0.09523 i \\
      &    & 0.37703 - 0.08911 i & 0.37706 - 0.08918 i \\
      &    &  &   \\
      & 0.4   & 0.49741 - 0.09756 i & 0.49741 - 0.09755 i \\
         &    & 0.47091 - 0.09584 i & 0.47091 - 0.09584 i \\
         &    & 0.47114 - 0.09586 i & 0.47115 - 0.09585 i \\
      &    & 0.38798 - 0.08978 i & 0.38798 - 0.08982 i \\
      &    &  &   \\
      3   & 0   & 0.67537 - 0.09650 i & 0.67537 - 0.09650 i \\
         &    & 0.65690 - 0.09562 i & 0.65690 - 0.09562 i \\
         &    & 0.65690 - 0.09562 i & 0.65690 -  0.09562 i \\
      &    & 0.59944 - 0.09270 i & 0.59944 - 0.09270 i \\
      &    &  &   \\
      & 0.2   & 0.67995 - 0.09671 i & 0.67995 - 0.09671 i \\
         &    & 0.66142 - 0.09583 i & 0.66142 - 0.09583 i \\
         &    & 0.66144 - 0.09583 i & 0.66144 - 0.09583 i \\
      &    & 0.60411 - 0.09294 i & 0.60411 - 0.09294 i \\
      &    &  &  \\
      & 0.4   & 0.69454 - 0.09730 i & 0.69455 - 0.09730 i \\
         &    & 0.67560 - 0.09643 i & 0.67570 - 0.09643 i \\
         &    & 0.67591 - 0.09644 i & 0.67593 - 0.09645 i \\
      &    & 0.61905 - 0.09362 i & 0.61906 - 0.09362 i  \\
      &    &  &  \\
   \hline
    \end{tabular}
    \caption{The table presents the fundamental ( $n=0$) QNM frequencies for scalar (row 1), axial electromagnetic (row 2), polar electromagnetic (row 3), and gravitational (row 4) perturbations of the EB black hole, computed via the sixth-order WKB approximation, the AIM (after 100 iterations).}
    \label{Table 1}
\end{table}

\begin{table}[ht!]
    \centering
    \begin{tabular}{|l|c|c|r|}
     \hline
       $n$ & $q_m[M]$ & WKB & AIM  \\
    \hline
      1   & 0   & 0.46385 - 0.29563 i & 0.46385 - 0.29560 i  \\
         &    & 0.43653 - 0.29073 i & 0.43654 - 0.29071 i  \\
         &    & 0.43653 - 0.29073 i & 0.43654 - 0.29071 i  \\
      &    & 0.34630 - 0.27348 i & 0.34671 - 0.27391 i  \\
      &    &  &  \\
         & 0.2   & 0.46729 - 0.29621 i & 0.46729 - 0.29618 i  \\
         &    & 0.43992 - 0.29134 i & 0.43994 - 0.29132 i  \\
         &    & 0.43993 - 0.29134 i & 0.43994 - 0.29132 i  \\
      &    & 0.35017 - 0.27407 i & 0.35047 - 0.27452 i  \\
      &    &  &   \\
      & 0.4   & 0.47829 - 0.29782 i & 0.47830 - 0.29779 i  \\
         &    & 0.45057 - 0.29302 i & 0.45082 - 0.29305 i  \\
         &    & 0.45081 - 0.29306 i & 0.45082 - 0.29305 i  \\
      &    & 0.36253 - 0.27592 i & 0.36261 - 0.27624 i  \\
      &    &  &  \\
      2   & 0   & 0.43039 - 0.50870 i & 0.43054 - 0.50856 i  \\
         &    & 0.40091 - 0.50173 i & 0.40119 - 0.50159 i  \\
         &    & 0.40091 - 0.50173 i & 0.40119 - 0.50159 i  \\
      &    & 0.29852 - 0.47756 i & 0.30105 - 0.47828 i \\
      &    &  &  \\
         & 0.2   & 0.43409 - 0.50952 i & 0.43424 - 0.50938 i  \\
         &    & 0.40458 - 0.50259 i & 0.40485 - 0.50246 i  \\
         &    & 0.40459 - 0.50260 i & 0.40485 - 0.50246 i  \\
      &    & 0.30334 - 0.47818 i & 0.30553 - 0.47906 i  \\
      &    &  &  \\
      & 0.4   & 0.44597 - 0.51172 i & 0.44609 - 0.51157 i  \\
         &    & 0.41617 - 0.50488 i & 0.41664 - 0.50481 i  \\
         &    & 0.41643 - 0.50493 i & 0.41664 - 0.50481 i  \\
      &    & 0.31864 - 0.48046 i & 0.31992 - 0.48120 i \\
      &    &  &  \\
\hline
    \end{tabular}
    \caption{The table presents the non fundamental QN frequencies for scalar (row 1), axial electromagnetic (row 2), polar electromagnetic (row 3), and gravitational (row 4) perturbations of the EB black hole, computed via the sixth-order WKB approximation, the AIM (after 40 iterations). Here we set $l=2$}
    \label{Table 2}
\end{table}

\begin{figure*}[t]\centering
\includegraphics[width=0.45\textwidth]{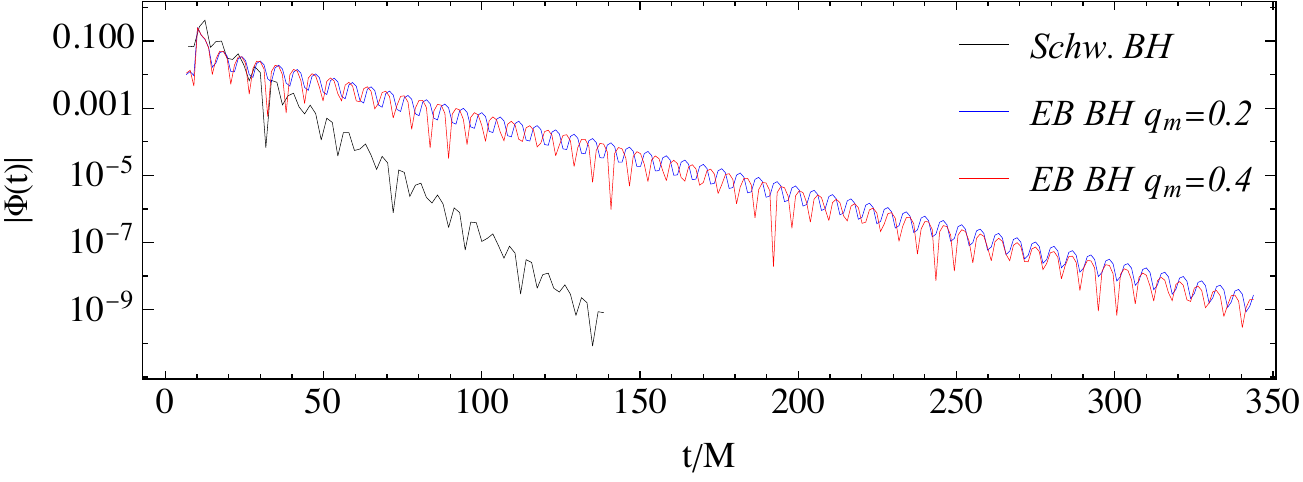}
\includegraphics[width=0.45\textwidth]{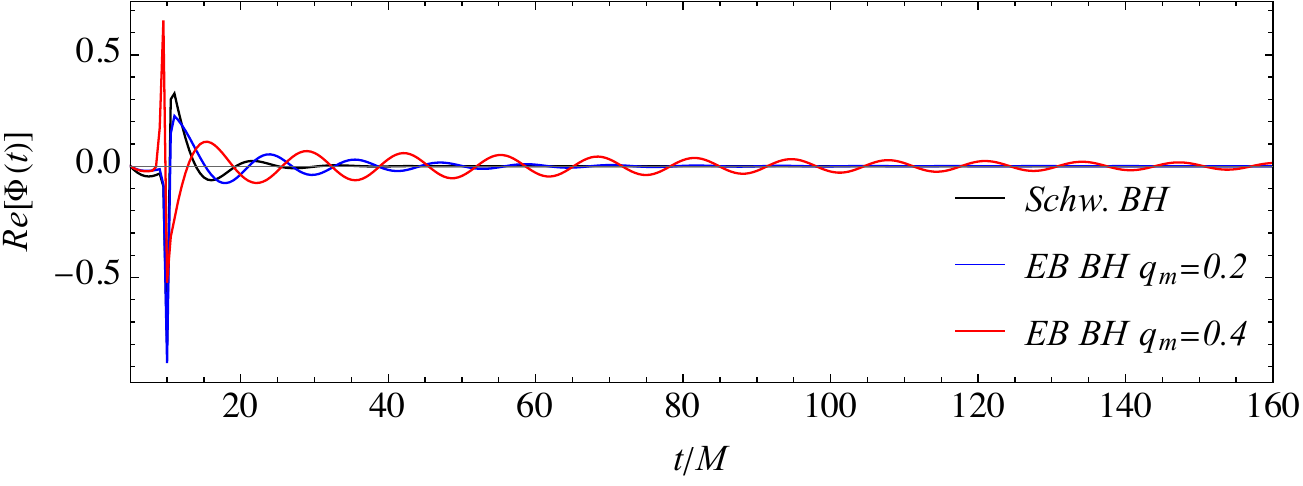}
\includegraphics[width=0.45\textwidth]{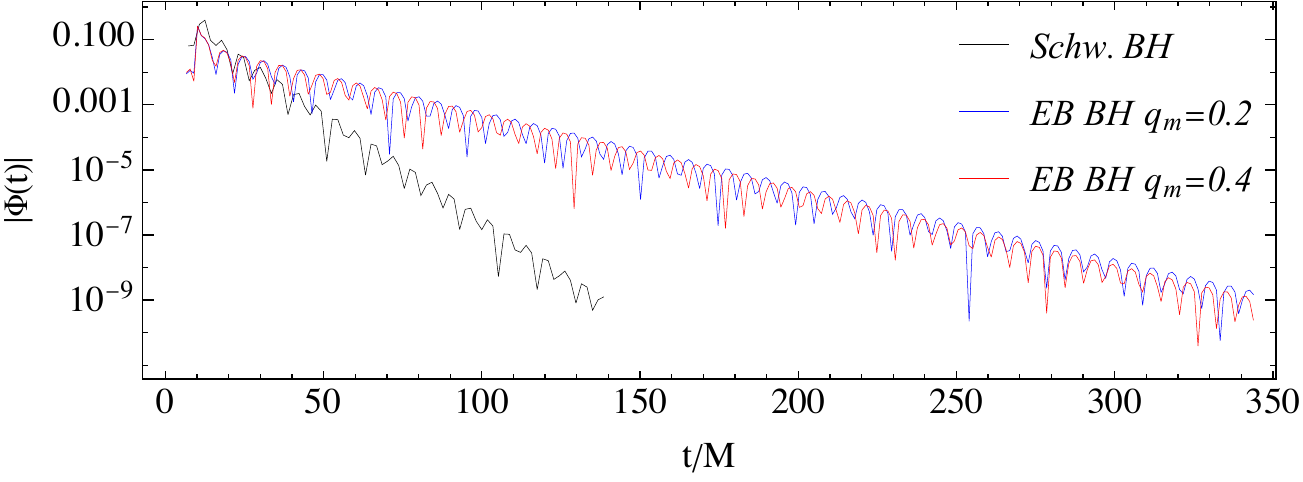}
\includegraphics[width=0.45\textwidth]{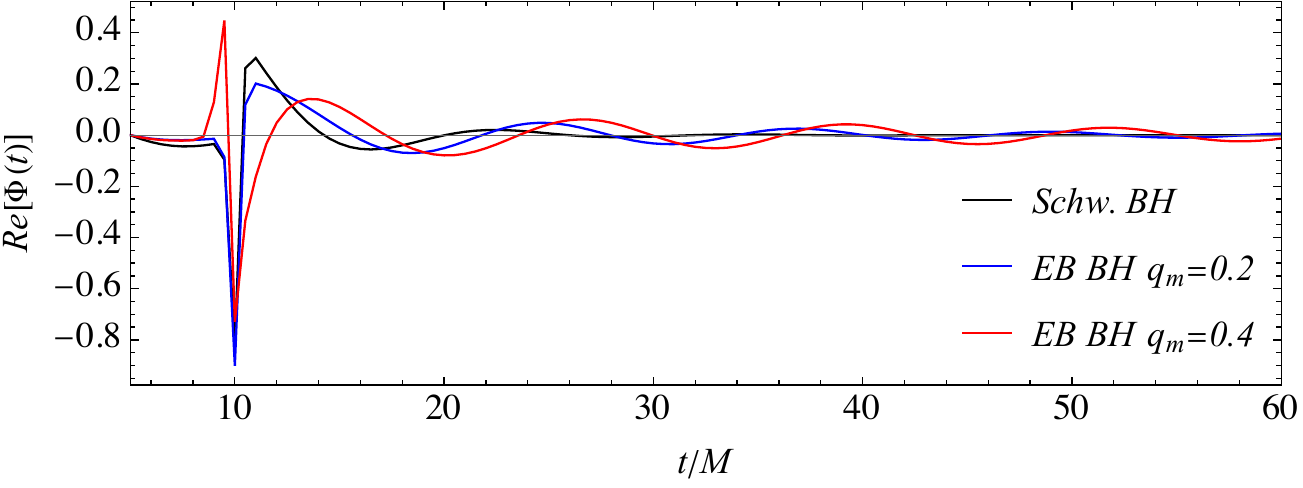}
\includegraphics[width=0.45\textwidth]{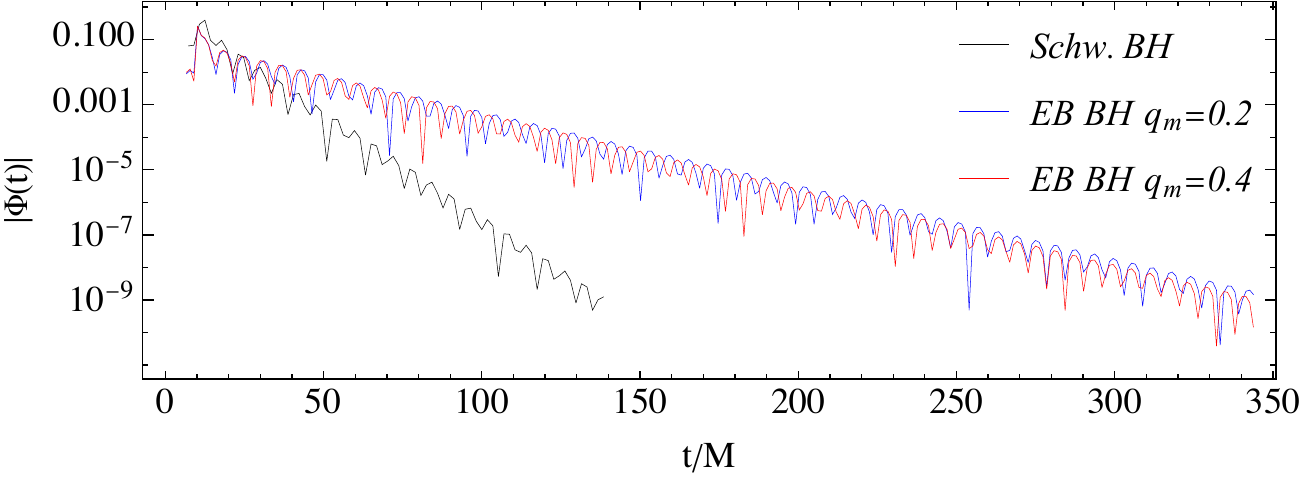}
\includegraphics[width=0.45\textwidth]{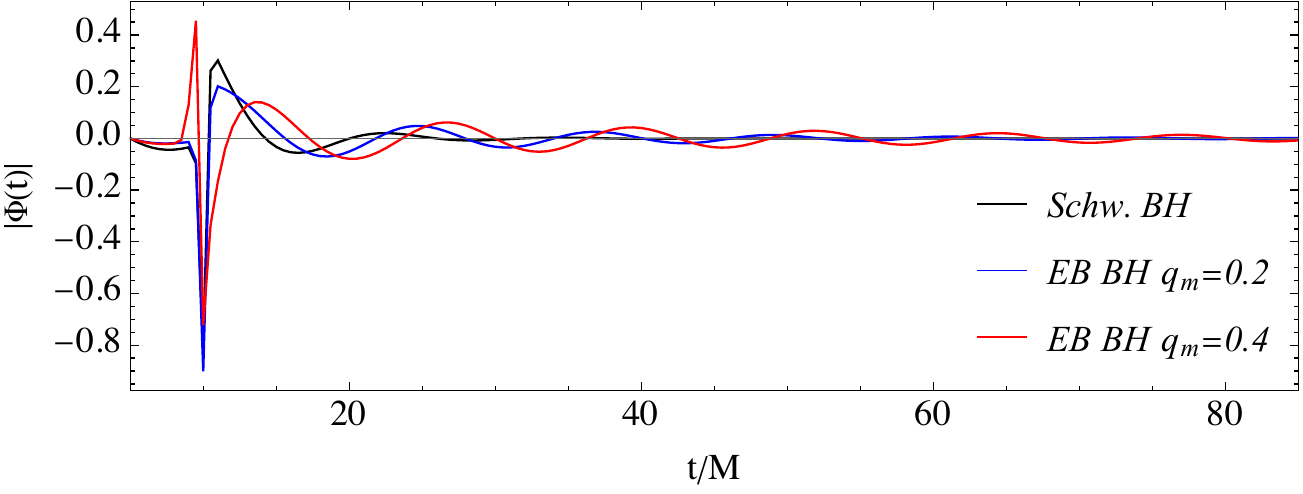}
\includegraphics[width=0.45\textwidth]{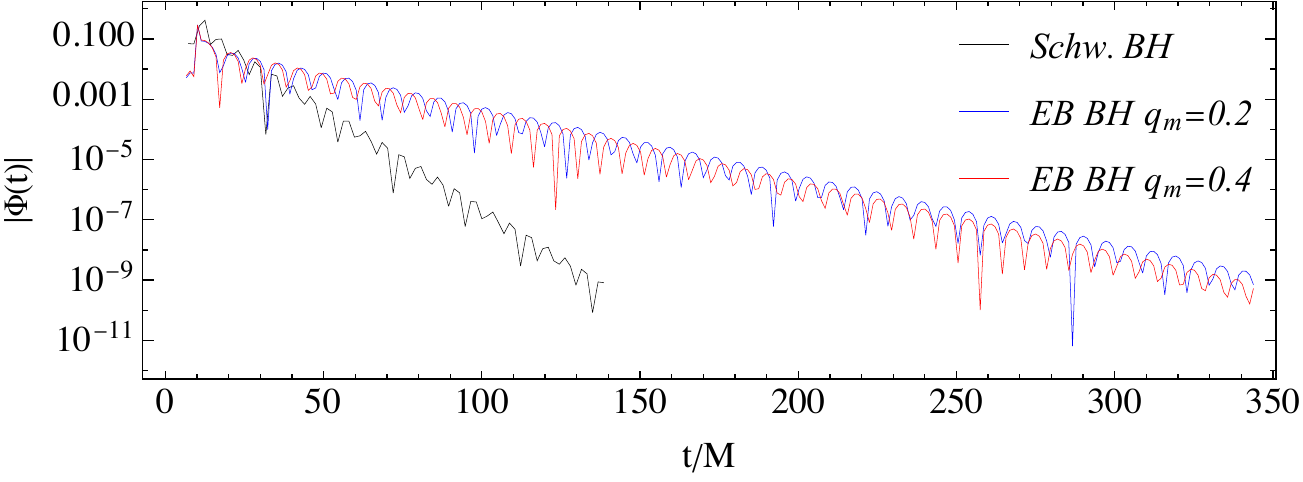}
\includegraphics[width=0.45\textwidth]{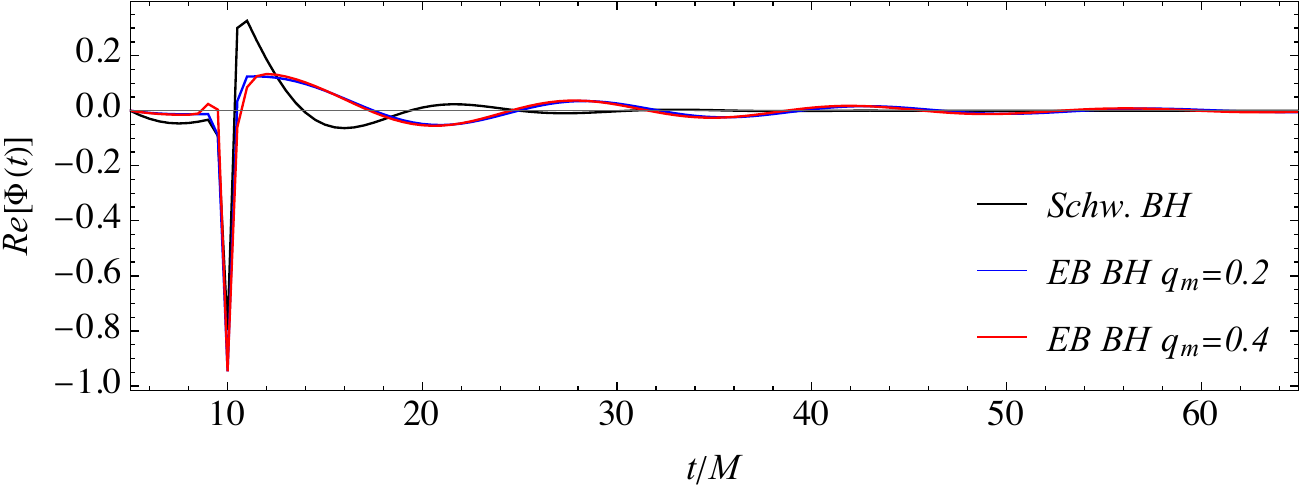}
\caption{The temporal evolution of the scalar field (the first row), the axial electromagnetic field (the second row), the polar electromagnetic perturbations (the third row), and the gravitational perturbations (the fourth row) is presented.}
    \label{fig. Phi}
\end{figure*}

\section{The temporal evolution of the all types of the perturbations}
In this part, we examine how the scalar, axial and polar electromagnetic, and gravitational perturbations of the  EB black hole within NLED evolve over time. To do so, we employ a characteristic integration scheme (\cite{Chirenti:2007mk}) based on light-cone coordinates, using the retarded time $du = dt - dx$ and the advanced time $ dv = dt + dx$, with initial conditions set on the two null surfaces $u = u_0$ and $v = v_0$. Subsequently, Eq.(\ref{eq.wave1}) can be rexpressed as:
\begin{eqnarray}
    -4\frac{\partial^2\Psi_s}{\partial u\partial v}=V_s(r(u,v))\Psi_s\,.
\end{eqnarray}
The numerical procedure employed to solve this equation is as follows:
\begin{eqnarray}
    \Psi_N=\left(\Psi_W+\Psi_E\right)\frac{16-\Delta^2V_C}{16+\Delta^2V_C}-\Psi_C\,,
\end{eqnarray}
in which we discretize the $(u, v)$ space using a uniform grid, where neighboring points are separated by a fixed interval $\Delta$ and the labels N, W, E,  C denote the individual grid points: $N\equiv(u,v)\,,\,W\equiv(u-\Delta,v)\,,\,E\equiv(u,v-\Delta)\,,\,C\equiv(u-\Delta,v-\Delta)$.

Our simulations are initialized with a Gaussian function of width $\sigma$
centered on $v_c$, following (\cite{UktamjonUktamov:2026dep}):
\begin{eqnarray}
    \Psi(u_0,v)=A\exp{\left[-\frac{(v-v_c)^2}{2\sigma^2}\right]}\,.
\end{eqnarray}
Finally, the temporal evolution of the scalar,axial and polar electromagnetic, and gravitational perturbations of the EB black hole for chosen values of the charge $q_m$ is illustrated in Fig. (\ref{fig. Phi}).

\section{Conclusion}

In this work, we have carried out a comprehensive analysis of scalar, electromagnetic, and gravitational perturbations of the EB  black hole arising in Einstein gravity coupled to NLED. Starting from the exact regular black hole solution, we first examined the spacetime geometry and demonstrated that, similarly to the Reissner-Nordström spacetime, the EB black hole possesses inner and outer horizons that merge at the extremal magnetic charge. The comparison with the Reissner-Nordström black hole showed that NLED modifies the horizon structure while preserving the existence of an extremal configuration.

We subsequently derived the master wave equations governing massless scalar, axial electromagnetic, polar electromagnetic, and gravitational perturbations. The corresponding effective potentials were obtained explicitly and analyzed in detail. Our results showed that all perturbations are characterized by positive potential barriers that vanish both at the event horizon and at spatial infinity, ensuring the appropriate asymptotic behavior required for QN oscillations. The effective potentials were found to increase with increasing magnetic charge, indicating that NLED significantly alters the scattering properties of perturbation fields. Moreover, among the perturbations considered, the scalar field exhibits the highest potential barrier. We also demonstrated that the axial and polar electromagnetic perturbations are governed by different effective potentials, reflecting the breaking of the isospectrality that is characteristic of Maxwell electrodynamics in linear theories. This distinction originates from the nonlinear electromagnetic coupling and represents one of the characteristic signatures of the EB spacetime.

The QN spectrum was then investigated using two independent approaches, namely the sixth-order WKB approximation and the AIM. The excellent agreement obtained between these methods for both the fundamental and higher overtones confirms the reliability and numerical stability of the calculated spectra over the considered range of magnetic charge. The obtained QN frequencies reveal that increasing the magnetic charge generally shifts the oscillation frequencies toward larger real values, while the imaginary parts undergo comparatively smaller variations. Consequently, NLED primarily modifies the oscillation frequencies of the perturbations while preserving their damping behavior, indicating that the EB black hole remains dynamically stable under scalar, electromagnetic, and gravitational perturbations. Furthermore, the systematic differences between the scalar, electromagnetic, and gravitational spectra demonstrate that each type of perturbation responds differently to the underlying nonlinear electromagnetic geometry.

To further verify these results, we investigated the temporal evolution of all perturbation fields using the characteristic integration method in double-null coordinates. The time-domain profiles exhibit the expected sequence of an initial transient stage followed by exponentially damped oscillations governed by the fundamental QNMs. The temporal evolution is fully consistent with the frequencies obtained from both the WKB and AIM, providing an independent confirmation of the QNM analysis. No unstable growing modes were observed for any perturbation considered, supplying additional evidence for the linear dynamical stability of the EB black hole.

Overall, our analysis demonstrates that NLED leaves clear imprints on both the effective potentials and the QN spectra of perturbation fields. The increase of the potential barriers with magnetic charge, the non-isospectral behavior of axial and polar electromagnetic perturbations, and the magnetic-charge dependence of the QN frequencies collectively distinguish the EB black hole from its linear electrodynamics counterparts. These results provide new insight into the dynamical properties of regular black holes supported by NLED and may prove useful for future investigations of GW ringdown signals, black hole spectroscopy, and observational tests aimed at distinguishing regular black hole geometries from conventional charged black holes.

\section*{Acknowledgements}

\section{Appendix}\label{app}
Variables for Eq.(\ref{eq.Field eq.}):
\begin{subequations}
    \begin{align}
&\alpha=\frac{f(r)\cosh^2{\frac{q_m^2}{2Mr}}}{1-\frac{q_m^2}{4Mr}\tanh{\frac{q_m^2}{2Mr}}}\,,\\
&\gamma=\Big(1+\frac{4q_m^2\left[q_m^2\left(-2+\cosh{\frac{q_m^2}{Mr}}\right)-5Mr\sinh{\frac{q_m^2}{Mr}}\right]}{r^2M\left(4M-\frac{q_m^2}{r}\tanh{\frac{q_m^2}{2Mr}}\right)}\Big)
    \end{align}
\end{subequations}

\allowdisplaybreaks


\bibliography{prd/main1}    
\end{document}